\documentclass[a4paper,10pt,eqno]{article}
\usepackage{theorem}
\usepackage{latexsym,amssymb,amsfonts,amsmath}
\usepackage{graphicx}
\newtheorem{thm}{Theorem}[section]
\newtheorem{lem}[thm]{Lemma}
\newtheorem{cor}[thm]{Corollary}
\newtheorem{pro}[thm]{Proposition}

\theoremheaderfont{\scshape}
\newcommand{\RM}{\mathbb{R}}
\newcommand{\ZM}{\mathbb{Z}}

\newcommand{\CM}{\mathbb{C}}

\newcommand{\ket}[1]{|#1 \rangle}

\title{{\Large {\bf A note on It\^o's formula for discrete-time quantum walk}}}
\author{
{\small Norio Konno}\\
{\scriptsize Department of Applied Mathematics, 
Faculty of Engineering, 
Yokohama National University}\\
{\scriptsize Hodogaya, Yokohama 240-8501, Japan}\\
{\scriptsize e-mail: konno@ynu.ac.jp, Tel.: +81-45-339-4205, Fax: +81-45-339-4205}\\
}
\date{\empty }
\begin{document}
\maketitle

\par\noindent
\begin{small}
\par\noindent
{\bf Abstract}. We present an It\^o's formula for the one-dimensional discrete-time quantum walk and give some examples including a Tanaka's formula by using the formula. Moreover we discuss integrals for the quantum walk.

\footnote[0]{
{\it Abbr. title:} It\^o's formula for discrete-time quantum walk
}
\end{small}

\setcounter{equation}{0}
\section{Introduction}
The It\^o formula for Brownian motion is famous for the stochastic calculus. It\^o's stochastic calculus is a very useful tool for mathematical finance, stochastic control and filtering problems, see Kunita \cite{Kunita2010}, for example. It\^o's formula for the random walk has also been investigated (see Fujita \cite{Fujita2002, Fujita2008}). Quantum walks were introduced as the quantum counterparts and they have been intensively studied for the last decade (\cite{Kempe2003, Kendon2007, Konno2008, VAndraca2008}). However it is not known that It\^o's formula for the discrete-time quantum walk. Therefore we consider an It\^o's formula for the walk in this note. The rest of the manuscript is organized as follows. Section 2 is devoted to the definition of the quantum walk we consider. In Section 3, we present an It\^o's formula for the one-dimensional discrete-time quantum walk and give some examples including a Tanaka's formula by using the formula. In the final section, we discuss integrals for the quantum walk.

\section{Definition of the walk}
In this section, we briefly give the definition of the two-state quantum walk on $\ZM$ considered here, where $\ZM$ is the set of integers. The discrete-time quantum walk is a quantum version of the classical random walk with an additional degree of freedom called chirality. The chirality takes values left and right, and it means the direction of the motion of the walker. At each time step, if the walker has the left chirality, it moves one step to the left, and if it has the right chirality, it moves one step to the right. In this note, we put
\begin{eqnarray*}
\ket{L} = 
\left[
\begin{array}{cc}
1 \\
0  
\end{array}
\right],
\qquad
\ket{R} = 
\left[
\begin{array}{cc}
0 \\
1  
\end{array}
\right],
\end{eqnarray*}
where $L$ and $R$ refer to the left and right chirality state, respectively.  

For the general setting, the time evolution of the walk is determined by a $2 \times 2$ unitary matrix, $U$, where
\begin{align*}
U =
\left[
\begin{array}{cc}
a & b \\
c & d
\end{array}
\right],
\end{align*}
with $a, b, c, d \in \mathbb C$ and $\CM$ is the set of complex numbers. The matrix $U$ rotates the chirality before the displacement, which defines the dynamics of the walk. To describe the evolution of our model, we divide $U$ into two matrices:
\begin{eqnarray*}
P_{-1} =
\left[
\begin{array}{cc}
a & b \\
0 & 0 
\end{array}
\right], 
\quad
P_{1} =
\left[
\begin{array}{cc}
0 & 0 \\
c & d 
\end{array}
\right],
\end{eqnarray*}
with $U = P_{-1} + P_1$. The important point is that $P_{-1}$ (resp. $P_1$) represents that the walker moves to the left (resp. right) at position $x$ at each time step. 

The {\it Hadamard walk} is determined by the Hadamard gate $U = H$:
\begin{eqnarray*}
H=\frac{1}{\sqrt2}
\left[
\begin{array}{cc}
1 & 1 \\
1 &-1 
\end{array}
\right].
\end{eqnarray*}
The walk is intensively investigated in the study of the quantum walk. 

In the present note, we take $\varphi = {}^T [\alpha, \beta]$ with $\alpha, \> \beta \in \CM$ and $|\alpha|^2 + |\beta|^2 = 1$ as the initial qubit state, where $T$ is the transpose operator.

Let $\Xi_{n} (l,m)$ denote the sum of all paths starting from the origin in the trajectory consisting of $l$ steps left and $m$ steps right at time $n$ with $n=l+m$. For example, 
\begin{align*}
\Xi_2 (1,1) &= P_1 P_{-1} + P_{-1} P_1, \\
\Xi_4 (2,2) &= P_1^2 P_{-1}^2 + P_{-1}^2 P_1^2 + P_1 P_{-1} P_1 P_{-1} + P_{-1} P_1 P_{-1} P_1 + P_{-1} P_1^2 P_{-1} + P_1 P_{-1}^2 P_1. 
\end{align*}
The probability that our quantum walker is in position $x \> (\in \ZM)$ at time $n \> (\in \{0,1, \ldots \})$ starting from the origin with $\varphi = {}^T [\alpha, \beta]$ with $\alpha, \> \beta \in \CM$ and $|\alpha|^2 + |\beta|^2 = 1$ is defined by 
\begin{align*}
P (X_{n} =x) = || \Xi_{n}(l, m) \> \varphi ||^2,
\end{align*}
where $n=l+m$ and $x=-l+m$. We define the probability amplitude of the quantum walk in position $x$ at time $n$ by 
\begin{equation*}
\Psi_{n}(x)=
\left[
\begin{array}{cc}
\Psi_{n}^{L}(x) \\
\Psi_{n}^{R}(x)
\end{array}
\right].
\label{eqn:quantum1}
\end{equation*}
Then we see that
\begin{align*}
P (X_{n} =x) = || \Psi_{n}(x) ||^2 = |\Psi_{n}^{L}(x)|^2 + |\Psi_{n}^{R}(x)|^2.
\end{align*}
From now on we consider the Fourier transform of $\Psi_{n}(x).$ By definition,
\begin{equation}
\Psi_{n+1}(x) = P \> \Psi_{n}(x+1) + Q \> \Psi_{n}(x-1).
\label{eqn:U1}
\end{equation}
Moreover, the Fourier transform of ${\Psi}_{n}^j(x) \> (j = L, \> R)$, that is, $\hat{\Psi}_{n}^{j}(\xi)$, is defined by
\begin{eqnarray*}
\hat{\Psi}_{n}^{j}(\xi) = \sum_{x \in \ZM} e^{i \xi x} \Psi_{n}^{j}(x).
\end{eqnarray*}
Thus we have
\begin{eqnarray*}
\Psi_{n}^{j}(x) = \int_{-\pi}^{\pi} \frac{d\xi}{2 \pi} e^{-i \xi x} \hat{\Psi}_{n}^{j}(\xi).
\end{eqnarray*}
Here we put
\begin{eqnarray*}
\hat{\Psi}_{n}(\xi) = 
\left[
\begin{array}{cc}
\hat{\Psi}_{n}^{L}(\xi) \\
\hat{\Psi}_{n}^{R}(\xi)
\end{array}
\right],
\end{eqnarray*}
then
\begin{align*} 
\hat{\Psi}_{n} (\xi) 
= \sum_{x \in \ZM} e^{i \xi x} \Psi_{n}(x), \qquad \Psi_{n}(x) 
= \int_{-\pi}^{\pi} \frac{d\xi}{2 \pi} e^{-i \xi x} \hat{\Psi}_{n}(\xi).
\end{align*} 
From Eq. (\ref{eqn:U1}), for $\xi \in [-\pi, \pi)$Cwe obtain

\begin{lem}
\label{lem:rinko}
For any $n=0,1,2, \ldots$,
\begin{equation*}
\hat{\Psi}_{n+1}(\xi) = U(\xi) \> \hat{\Psi}_{n}(\xi).
\label{eqn:woods}
\end{equation*}
Here $U(\xi)$ is given by
\begin{equation}
\label{eqn:Uxi}
U(\xi)= 
e^{-i\xi} \> P_{-1} + e^{i\xi} \> P_1 
=
\left[
\begin{array}{cc}
e^{-i\xi} & 0 \\
0 &  e^{i\xi}
\end{array}
\right]
\> U.
\end{equation}
\end{lem}
We should remark that
\begin{equation*}
\hat{\Psi}_{0}(\xi)=
\left[
\begin{array}{cc}
\alpha  \\
\beta
\end{array}
\right],
\end{equation*}
where $\alpha, \beta \in \CM, |\alpha|^2+|\beta|^2=1$. By Lemma {\rmfamily \ref{eqn:U1}}, we have
\begin{equation*}
\hat{\Psi}_{n}(\xi) = U(\xi)^n \> \hat{\Psi}_{0}(\xi).
\end{equation*}
It is important for the study of the quantum walk to compute $U(\xi)^n,$ since
\begin{align*}
P (X_{n} =x) = || \Psi_{n}(x) ||^2 =  \int_{-\pi}^{\pi} \frac{d\xi'}{2 \pi} e^{i \xi' x} \left( U(\xi')^n \>\hat{\Psi}_{0}(\xi') \right)^{\ast} \> \int_{-\pi}^{\pi} \frac{d\xi}{2 \pi} e^{-i \xi x}  U(\xi)^n \> \hat{\Psi}_{0}(\xi),
\end{align*}
where $\ast$ means the adjoint operator.

\section{It\^o's formula for quantum walk}
Let $B_n = \{-n, -(n-1), \ldots , n-1,n \}$ and $\Omega_n = B_n ^{n+1} = \{-n, -(n-1), \ldots , n-1,n \}^{n+1}.$ From now on, we will consider quantum walks on the path space $\Omega_n$. To do so, we let $w_n = (w_n(0)=0, w_n (1), w_n (2), \ldots , w_n (n)) \in \Omega_n$. Next we introduce 
\begin{align*}
v_n = (v_n (1), v_n (2), \ldots , v_n (n)) = 
(w_n (1), w_n (2) - w_n (1), \ldots , w_n (n)-w_n (n-1)),
\end{align*}
and
\begin{align*}
u_n = (u_n (1), u_n (2), \ldots , u_n (n)) = 
(I_{\{1\}} (v_n (1)), I_{\{1\}} (v_n (2)), \ldots , I_{\{1\}} (v_n (n))),
\end{align*}
where $I_A (x)$ denotes the indicator function for a set $A$. Noting that $w_n(m+1) - w_n(m) \in \{ 1, -1 \}$, a direct computation gives
\begin{pro}
\label{trib}
Let $f$ be a function on $\ZM$ with values in $\CM$. For any $m \in \{0,1, \ldots , n-1\},$
\par\noindent
(1) 
\begin{align*} 
f(w_n(m+1)) 
&- f(w_n(m)) 
\\ 
&= \frac{1}{2} \left\{ f(w_n(m)+1) - f(w_n(m)-1) \right\} 
\left( w_n(m+1) - w_n(m) \right)
\\
&+ \frac{1}{2} \left\{ f(w_n(m)+1) - 2 f(w_n(m)) + f(w_n(m)-1) \right\}.
\end{align*} 
\par\noindent
(2) 
\begin{align*} 
f(w_n(n)) 
&- f(w_n(0)) 
\\ 
&= \frac{1}{2} \sum_{m=0}^{n-1} \left\{ f(w_n(m)+1) - f(w_n(m)-1) \right\} 
\left( w_n(m+1) - w_n(m) \right)
\\
&+ \frac{1}{2} \sum_{m=0}^{n-1} \left\{ f(w_n(m)+1) - 2 f(w_n(m)) + f(w_n(m)-1) \right\}.
\end{align*} 
\end{pro}
In fact, this result holds for the corresponding random walk. Let $\{ Y_n ; n=0,1,2, \ldots \}$ denote a simple symmetric random walk, that is, $Y_0 =0, \>\> Y_n = \xi_1 + \xi_2 + \cdots + \xi_n$, where $\xi_1, \xi_2, \ldots $ are independent and identically distributed with $P(\xi_1=1) = P(\xi_1=-1)=1/2$. Let $\RM$ be the set of real numbers. Then for any $\RM$-valued function $f$ on $\ZM$ and any $m \in \{0,1, \ldots , n-1\},$ it holds that
\begin{align*} 
f(Y_{m+1}) - f(Y_m) 
&= \frac{1}{2} \left\{ f(Y_m +1) - f(Y_m -1) \right\} \left( Y_{m+1} - Y_m \right)
\\
&+ \frac{1}{2} \left\{ f(Y_m+1) - 2 f(Y_m) + f(Y_m-1) \right\}.
\end{align*} 
This equation is called a discrete It\^o's formula (see Fujita \cite{Fujita2002}, Fujita and Kawanishi \cite{FujitaKawanishi2008}, for example). Therefore part (1) in Proposition {\rmfamily \ref{trib}} can be considered as an {\it It\^o's formula of the quantum walk}. Moreover 
\begin{align*} 
f(Y_{n}) - f(Y_0) 
&= \frac{1}{2} \sum_{m=0}^{n-1} \left\{ f(Y_m +1) - f(Y_m -1) \right\} \left( Y_{m+1} - Y_m \right)
\\
&+ \frac{1}{2} \sum_{m=0}^{n-1} \left\{ f(Y_m+1) - 2 f(Y_m) + f(Y_m-1) \right\}.
\end{align*} 
This is a Doob-Meyer's decomposition of the random walk, see \cite{Fujita2008, FujitaKawanishi2008}, for example. The first term of RHS is a martingale. We put $k= u_n (n) \> 2^{n-1} + u_n (n-1) \> 2^{n-2} + \cdots + u_n (2) \> 2^{1} + u_n (1) \> 2^{0}.$ To use this, we let
\begin{align*} 
P_{w_n^{(k)}} =P_{v_n^{(k)}(n)} \cdots P_{v_n^{(k)}(2)} P_{v_n^{(k)}(1)}. 
\end{align*} 
From Proposition {\rmfamily \ref{trib}}, we immediately obtain
\begin{thm}
\label{akina}
Let $f$ be a function on $\ZM$ with values in $\CM$. For any $m \in \{0,1, \ldots , n-1\},$
\par\noindent
(1) 
\begin{align*} 
&
\sum_{k=0}^{2^n-1} \left\{ f(w_n^{(k)}(m+1)) - f(w_n^{(k)}(m)) \right\} P_{w_n^{(k)}}
\\ 
&= \frac{1}{2} \sum_{k=0}^{2^n-1} \left\{ f(w_n^{(k)}(m)+1) - f(w_n^{(k)}(m)-1) \right\} 
\left( w_n^{(k)}(m+1) - w_n^{(k)}(m) \right) P_{w_n^{(k)}}
\\
&+ \frac{1}{2} \sum_{k=0}^{2^n-1} \left\{ f(w_n^{(k)}(m)+1) - 2 f(w_n^{(k)}(m)) + f(w_n^{(k)} (m)-1) \right\} P_{w_n^{(k)}}.
\end{align*} 
\par\noindent
(2) 
\begin{align*} 
& \sum_{k=0}^{2^n-1} \left\{ f(w_n^{(k)}(n)) - f(w_n^{(k)}(0)) \right\} P_{w_n^{(k)}}
\\ 
&= \frac{1}{2} \sum_{k=0}^{2^n-1} \sum_{m=0}^{n-1} \left\{ f(w_n^{(k)}(m)+1) - f(w_n^{(k)}(m)-1) \right\} 
\left( w_n^{(k)}(m+1) - w_n^{(k)}(m) \right) P_{w_n^{(k)}}
\\
&+ \frac{1}{2} \sum_{k=0}^{2^n-1} \sum_{m=0}^{n-1} \left\{ f(w_n^{(k)}(m)+1) - 2 f(w_n^{(k)}(m)) + f(w_n^{(k)}(m)-1) \right\} P_{w_n^{(k)}}.
\end{align*} 
\end{thm}
For example, we can check part (2) for $n=2$:
\begin{align*} 
LHS 
&= (f(-2) - f(0)) P_{-1}^2 + (f(2) - f(0)) P_{1}^2
\\
&= f(-2) P_{-1}^2 - f(0) \left( P_{-1}^2+P_{1}^2 \right) + f(2) P_{1}^2.
\end{align*} 
On the other hand, 
\begin{align*} 
RHS 
&= \frac{1}{2} \left( f(1) - f(-1) \right) \left\{ - \left(P_{-1}^2+ P_{1} P_{-1} \right) + \left( P_{-1} P_{1}+P_{1}^2 \right) \right\}
\\
&+ \frac{1}{2} \left( f(0) - f(-2) \right) \left( -P_{-1}^2 + P_{1} P_{-1} \right)
\\
&+ \frac{1}{2} \left( f(2) - f(0) \right) \left( -P_{-1} P_{1} + P_{1}^2 \right)
\\
&+ \frac{1}{2} \left( f(1) - 2f(0) + f(-1) \right) \left( P_{-1}+P_{1} \right)^2
\\
&+ \frac{1}{2} \left( f(0) - 2f(-1) + f(-2) \right) \left( P_{-1}^2+ P_{1} P_{-1} \right)
\\
&+ \frac{1}{2} \left( f(2) - 2f(1) + f(0) \right) \left( P_{-1} P_{1}+P_{1}^2 \right)
\\
&= f(-2) P_{-1}^2 - f(0) \left( P_{-1}^2+P_{1}^2 \right) + f(2) P_{1}^2.
\end{align*} 
When we consider $P_{-1} \to p \in [0,1]$ and $P_{1} \to q \in [0,1]$ with $p+q = 1$, Theorem {\rmfamily \ref{akina}} becomes the corresponding result for a simple random walk. 

We should remark that if we take $f(x)=x$ in part (2) with $w_n^{(k)}(0)=0$ for any $k$, we have the following trivial relation.
\begin{align*} 
LHS = \sum_{k=0}^{2^n-1} w_n^{(k)}(n) P_{w_n^{(k)}}.
\end{align*} 
On the other hand,
\begin{align*} 
RHS 
= \frac{1}{2} \sum_{k=0}^{2^n-1} \sum_{m=0}^{n-1} 2 
\left( w_n^{(k)}(m+1) - w_n^{(k)}(m) \right) P_{w_n^{(k)}} + 0 =
\sum_{k=0}^{2^n-1} w_n^{(k)}(n) P_{w_n^{(k)}}.
\end{align*} 

\par
Next we consider a {\it Tanaka's formula for the quantum walk}. Put $ \mathrm{sgn} (x) = 0 \> (x=0), \> =1 \> (x>0), \> =-1 \> (x<0)$. If we take $f(x)=|x|$ in part (2) with $w_n^{(k)}(0)=0$ for any $k$, then we obtain a Tanaka's formula as follows.
\begin{cor}
\begin{align*} 
& \sum_{k=0}^{2^n-1} \left| w_n^{(k)}(n) \right| P_{w_n^{(k)}}
\\ 
&= \sum_{k=0}^{2^n-1} \sum_{m=0}^{n-1} \mathrm{sgn} (w_n^{(k)}(m)) \left( w_n^{(k)}(m+1) - w_n^{(k)}(m) \right) P_{w_n^{(k)}} + \sum_{k=0}^{2^n-1} \sum_{m=0}^{n-1} I_{\{0\}}(w_n^{(k)}(m)) P_{w_n^{(k)}}.
\end{align*}
\end{cor} 
Here we used $(f(x+1)-f(x-1))/2= \mathrm{sgn}(x)$ and $(f(x+1)-2 f(x) + f(x-1))/2=I_{\{0\}}(x)$. The second term of RHS corresponds the ``local time" at the origin of the quantum walk. The author \cite{KonnoST} computed some sojourn times of the Hadamard walk in one dimension. The formula would be useful for computing local time at the origin. Another expression of Tanaka's formula for the quantum walk can be considered for $f(x)= \lceil x \rceil = \max (x-1,-x)$ as in the case of random walk (see Fujita \cite{Fujita2008}). Our case is $f(x) = |x| = \max (x, -x).$

We take $f(x)= e^{i \xi x}$ in part (2) with $w_n^{(k)}(0)=0$ for any $k$. Note that 
\begin{align*} 
\sum_{k=0}^{2^n-1} e^{i \xi w_n^{(k)}(n)} P_{w_n^{(k)}} = U (\xi)^n, \quad \text{and} \quad \sum_{k=0}^{2^n-1} e^{-i \xi w_n^{(k)}(0)} P_{w_n^{(k)}} = \left(P_{-1} + P_{1} \right)^n = U^n,
\end{align*}
since Eq. (\ref{eqn:Uxi}) gives $U (\xi)^n = \left(e^{-i \xi} P_{-1} + e^{i \xi} P_{1} \right)^n.$ Then we have
\begin{cor}
\begin{align*} 
U (\xi)^n
&= U^n
+ i \sin \xi \sum_{k=0}^{2^n-1} \sum_{m=0}^{n-1} e^{i \xi w_n^{(k)}(m)} \left( w_n^{(k)}(m+1) - w_n^{(k)}(m) \right) P_{w_n^{(k)}} 
\\
&+ \left( \cos \xi -1 \right) \sum_{k=0}^{2^n-1} \sum_{m=0}^{n-1} e^{i \xi w_n^{(k)}(m)} P_{w_n^{(k)}}.
\end{align*}
\end{cor} 
As we mentioned in the last part of the previous section, $U (\xi)^n$ is an important quantity for analysis of the quantum walk.

\section{Discussion}
Very recently, Gudder and Sorkin considered an integral for the quantum process, see \cite{Gudder2011a, Gudder2011b, GudderSorkin2011}, for example. Motivated by their study, this section treats integrals of quantum walks. For $w_n \in \Omega_n$, we let $f(w_n) = f(w_n (0), w_n (1), \ldots, w_n (n))$. In the previous section, we considered the following type of the integral on the path space $\Omega_n$:
\begin{align*}
\sigma_n (f) = \sum_{k=0}^{2^n-1} f(w_n^{(k)}) P_{w_n^{(k)}}.
\end{align*} 
For the classical random walk case, $\sigma_n (\cdot)$ becomes an expectation on the path space $\Omega_n$. Here we present a $2^n \times 2^n$ decoherence matrix: for $k, k' \in \{0,1, \ldots ,2^n-1\}$, 
\begin{align*}
D_n (w_n^{(k)},w_n^{(k')}) = \left\langle P_{w_n^{(k)}} \varphi, P_{w_n^{(k')}} \varphi \right\rangle.
\end{align*}
By using the matrix, if we consider the following type of the integral:
\begin{align*}
\int f d \mu_n 
&= \sum_{k=0}^{2^n-1} \sum_{k'=0}^{2^n-1} \min \left[ f(w_n^{(k)}), f(w_n^{(k')}) \right] D_n (w_n^{(k)}, w_n^{(k')})
\\
&= \sum_{k=0}^{2^n-1} \sum_{k'=0}^{2^n-1} \min \left[ f(w_n^{(k)}), f(w_n^{(k')}) \right] \left\langle P_{w_n^{(k)}} \varphi, P_{w_n^{(k')}} \varphi \right\rangle,
\end{align*} 
we find out a relation between $\int f d \mu_n$ and $\sigma_n (f)$ in the following way. If we let $f(w_n)= I_{A}(w_n)$ for a set $A \> (\subset \Omega_n)$, then noting 
\begin{align*}
\min \{ I_{A}(w_n), I_{A}(w'_n) \} = I_{A}(w_n) I_{A}(w'_n),
\end{align*}
we have
\begin{align*}
\int I_{A} d \mu_n 
&= \sum_{k=0}^{2^n-1} \sum_{k'=0}^{2^n-1} I_{A} (w_n^{(k)}) I_{A} (w_n^{(k')}) \left\langle P_{w_n^{(k)}} \varphi, P_{w_n^{(k')}} \varphi \right\rangle 
\\
&= \sum_{k=0}^{2^n-1} \sum_{k'=0}^{2^n-1} \left\langle I_{A} (w_n^{(k)}) P_{w_n^{(k)}} \varphi, I_{A} (w_n^{(k')}) P_{w_n^{(k')}} \varphi \right\rangle 
\\
&= \left\langle \sigma_n (I_{A}) \varphi , \sigma_n (I_{A}) \varphi \right\rangle . 
\end{align*} 
Moreover when we put $f(w_n)= I_{B_0 \times B_1 \times \cdots \times B_{n-1} \times \{x\}}(w_n) = I_{\{x\}} (w_n(n)) \> (x \in B_n)$, where $B_n = \{ -n, - (n-1), \ldots, n-1, n \}$, then we see that the probability distribution of the quantum walk can be written by using $\sigma_n (f)$ with $f=I_{B_0 \times B_1 \times \cdots \times B_{n-1} \times \{x\}}$:
\begin{align*}
P (X_{n} =x) = \left\langle \sigma_n (I_{B_0 \times B_1 \times \cdots \times B_{n-1} \times \{x\}}) \varphi , \sigma_n (I_{B_0 \times B_1 \times \cdots \times B_{n-1} \times \{x\}}) \varphi \right\rangle = || \sigma_n (I_{B_0 \times B_1 \times \cdots \times B_{n-1} \times \{x\}}) \varphi ||^2,
\end{align*}

Our final remark is that to clarify a connection between the noncommutative It\^o's formula based on quantum stochastic calculus due to Hudson and Parthasarathy (see \cite{Biane2010, HP1984}) and our formula would be one of the interesting future problems.

\par
\
\par\noindent
{\bf Acknowledgment.} This work was partially supported by the Grant-in-Aid for Scientific Research (C) of Japan Society for the Promotion of Science (Grant No. 21540118).
\par
\
\par

\begin{small}
\bibliographystyle{jplain}

\end{small}

\end{document}